\documentclass[twocolumn,english]{article}
\usepackage[T1]{fontenc}
\usepackage[latin9]{inputenc}
\usepackage{geometry}
\usepackage{textcomp}
\usepackage{graphicx}
\usepackage{babel}
\usepackage{xcolor, soul}

\begin{document}
\date{}

\twocolumn[
\begin{@twocolumnfalse} 

\title{\textbf{A viable laser driver for a user plasma accelerator}}

\maketitle
\author {L.A.Gizzi\textsuperscript{a}, }
\author {P. Koester\textsuperscript{a},}
\author {L. Labate\textsuperscript{a},}
\author {F. Mathieu\textsuperscript{b},}
\author {Z. Mazzotta\textsuperscript{b},}
\author {G. Toci\textsuperscript{c},}
\author {M. Vannini\textsuperscript{c},}

$\,$

{\textsuperscript{a}\textit{Intense Laser Irradiation Laboratory, Istituto Nazionale di Ottica, CNR, Pisa, Italy.}}

{\textsuperscript{b}\textit{Ecole Polytechnique, CNRS, Palaiseau, France }}

{\textsuperscript{c}\textit{Istituto Nazionale di Ottica, CNR, Firenze, Italy.}}

$\,$

\begin{abstract}

The construction of a novel user facility employing laser-driven plasma acceleration with superior beam quality will require an industrial grade, high repetition rate petawatt laser driver which is beyond existing technology. However, with the ongoing fast development of chirped pulse amplification and high average power laser technology, options can be identified depending on the envisioned laser-plasma acceleration scheme and on the time scale for construction. Here we discuss laser requirements for the EuPRAXIA infrastructure design and identify a suitable laser concepts that is likely to fulfil such requirements with a moderate development of existing technologies.

$\,$

$\,$
\end{abstract}
\end{@twocolumnfalse}
]

\section{Introduction}

Laser-driven plasma accelerators for user applications will require high repetition rate laser drivers capable of petawatt peak power and kW average laser power. The worldwide scenario for this kind of technology is rapidly evolving and new concepts are emerging with the promise of addressing medium and long term objectives of laser-plasma acceleration \cite{PhysRevLett.43.267} and future plasma-based particle colliders. However, scaling the technology of existing high peak power lasers based on Chirped Pulse Amplification (CPA) \cite{STRICKLAND1985447} to higher average power remains challenging. Pulsed high energy solid state lasers have demonstrated continuous operation at energies of approximately 100 J and repetition rates up to 10 Hz (e.g., RAL's DiPOLE100  \cite{Mason:17}       and the LLNL  HAPLS\cite{Sistrunk:17}). Solid-state lasers with peak power in the petawatt class \cite{danson_hillier_hopps_neely_2015} and pulse energies exceeding 10 J have reached an average power of tens of watts (BELLA \cite{7934119}), with HAPLS aiming at 300 W. In order to address user-level requirements, average power of ultrafast lasers will have to increase by one or two orders of magnitude.

In this scenario, the EuPRAXIA infrastructure design study \cite{Walker2017} aims at delivering a full concept for a compact, user-oriented, plasma accelerator with superior beam quality to enable free electron laser operation in the X-ray range and other developments of advanced compact radiation sources.
In its current configuration, the infrastructure relies on three main laser systems to drive plasma acceleration in different configurations, including the 150 MeV injector, the 1 GeV injector/accelerator and 5 GeV accelerator stage. 

\begin{figure} [ht]
\vspace{-2mm}\centering{}
\includegraphics[scale=0.2]{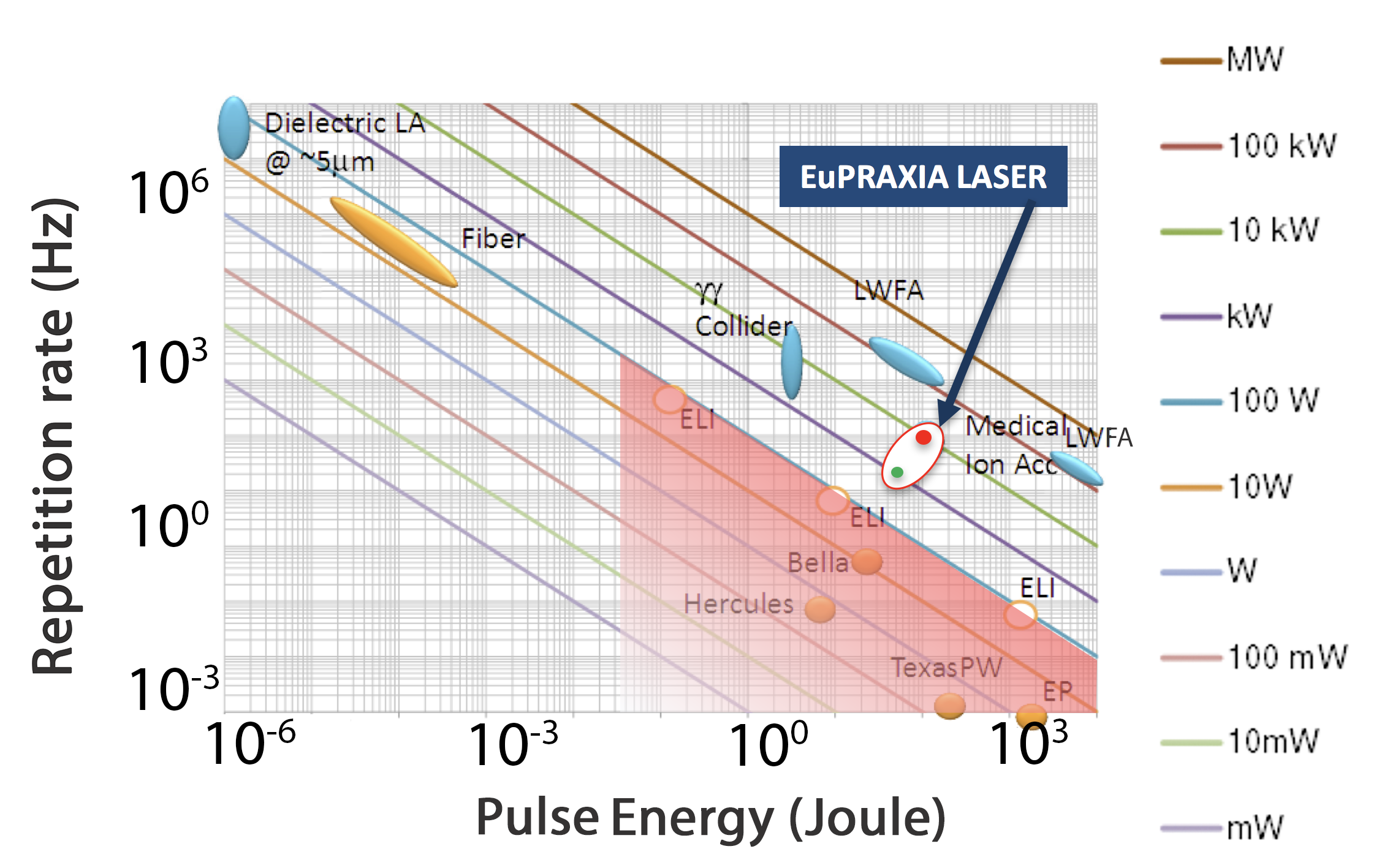}\vspace{-3mm}
\caption{The main EuPRAXIA laser specifications for the minimum mandatory configuration (green dot) and the best effort specifications (red dot).
\label{fig:scenario}\vspace{-2mm}}
\end{figure}

A pulse duration as short as 30 fs is required for the first two configurations with an energy per pulse as high as 100 J for the 5 GeV accelerator configuration. A repetition rate ranging from a minimum of 20 Hz up to 100 Hz is envisioned to fulfil user operation.
To achieve its goals the infrastructure will therefore require a laser driver ranging from 1 kW to 10 kW average laser power and petawatt peak power.
As shown in Figure \ref{fig:scenario}, such specifications are beyond current demonstrated performances but, as discussed here, they may be reached with incremental developments of current technology, although full accomplishment may require step changes based on novel technologies. 

\begin{table}
\centering{}
\includegraphics[scale=0.30]{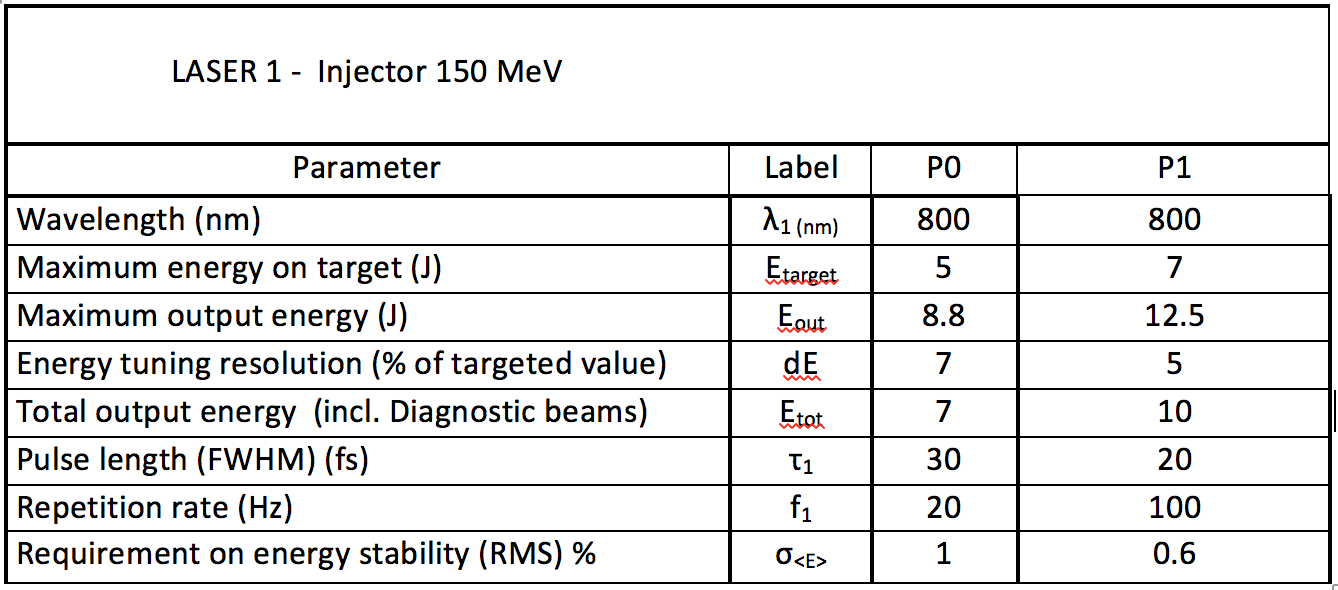}
\includegraphics[scale=0.30]{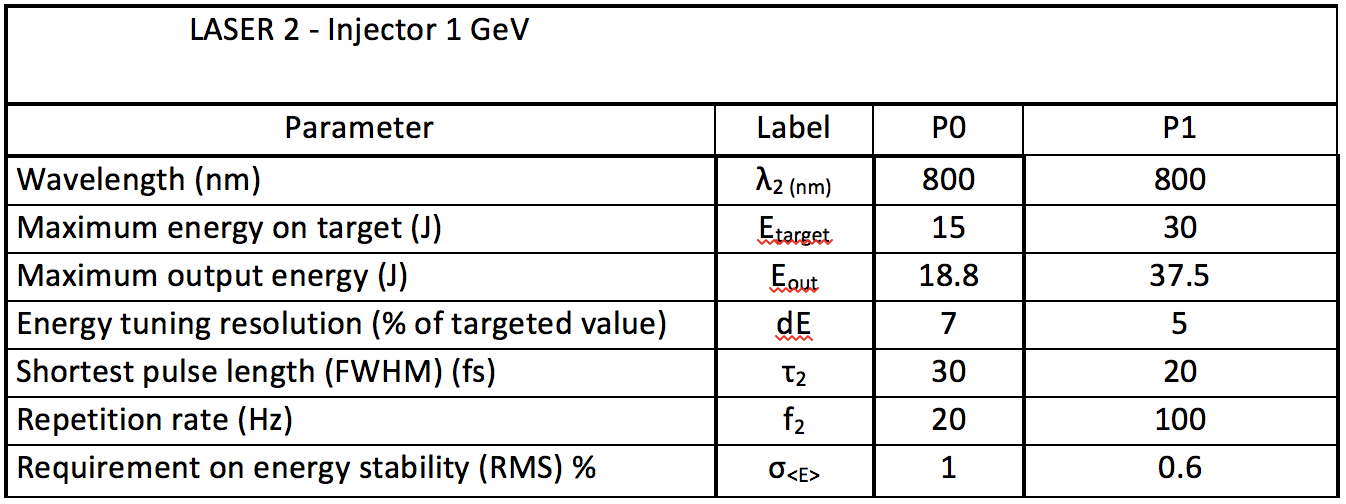}
\includegraphics[scale=0.29]{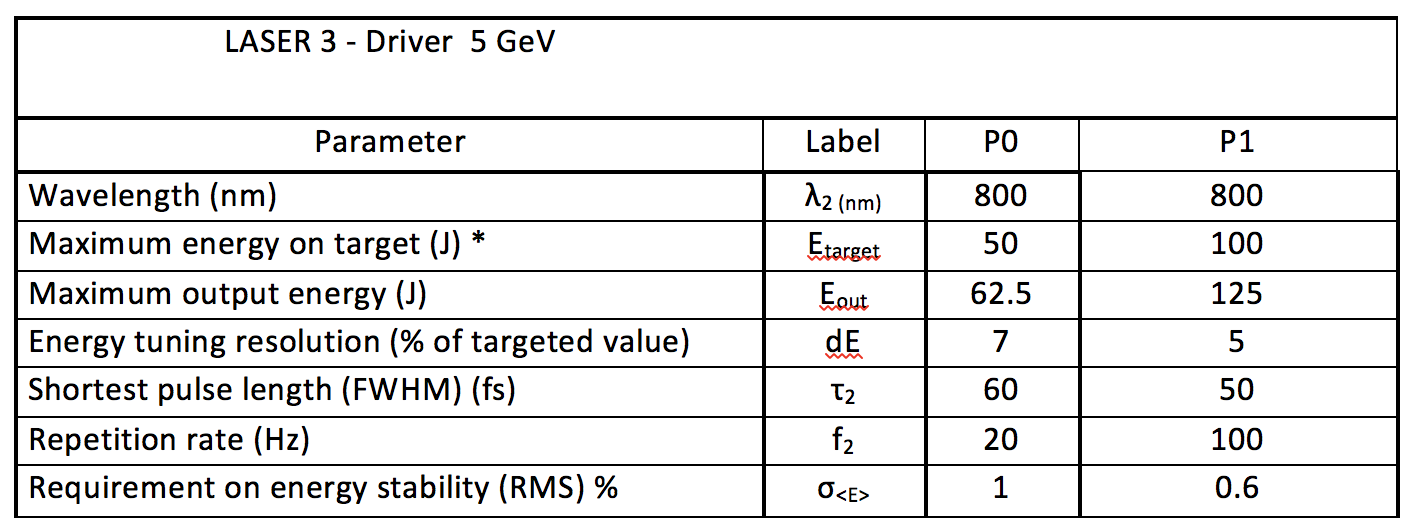}

\caption{Main requirements of the laser drivers for the injector at 150 MeV (top),  the injector at 1 GeV and (middle) and the accelerator at 5 GeV (bottom),  and for the mandatory (P0) and best effort (P1) performance levels.
\label{fig:laser_specs}}
\end{table}

Several technologies are developing which aim at high repetition rate, higher average power levels and energy-efficient configurations.  Fibre laser technology is currently offering the best wall-plug efficiency for a laser, now exceeding 50\% in CW mode, and solutions based on the coherent combination of a very large number of fiber amplifiers is being developed \cite{LeDortz:17, Klenke:17}. Such a technology is particularly suited and cost-competitive for applications requiring low energy per pulse and higher repetition rate, as technology is really optimised for repetition rate of 10 kHz and above.

Direct Chirped Pulse Amplification (DCPA) using lasing media that can be pumped directly with diodes may offer an additional alternative solution as it could enable higher efficiency and higher rep-rate. DCPA concepts currently under development, including Yb:YAG and Yb:CaF provide effective solution for 100 fs or longer pulses, while for shorter pulses like those envisioned for EuPRAXIA, new amplifying media are needed capable of a larger bandwidth. Recently, big aperture Tm:YLF  (BAT) concept has been proposed \cite{Haefnereaac2017} that offers significant lifetime advantage over Yb doped materials for multi-pulse extraction and becomes very efficient for kHz repetition rates and above. Moreover, the BAT architecture may allow amplification of shorter pulse duration than Yb based systems and  would be more efficient for wakefield excitation due to the longer plasma wavelength of $\lambda \approx 1.9\mu m$. 

The EuPRAXIA laser design focuses on a relatively short time-scale for construction of the infrastructure. In view of this, the laser design relies primarily on the most established laser technologies currently available, namely those that are likely to be scaled to the required specifications starting from existing prototypes. Titanium Sapphire technology pumped by 0.5 $\mu$m, diode-pumped solid state (DPSSL) lasers, provides a relatively safe ground, with major industrial endeavour in place, on which the required developments can also build. As discussed here, technologies addressing some of the most critical aspects related to the high average power operation, primarily pump sources, but also cooling strategies for the amplifying media and for the diffraction gratings, are reaching a readiness level  that can make it reasonably possible their deployment in the required time scale.

\section{Preliminary laser design}
\begin{figure}
\centering{}\includegraphics[scale=0.22]{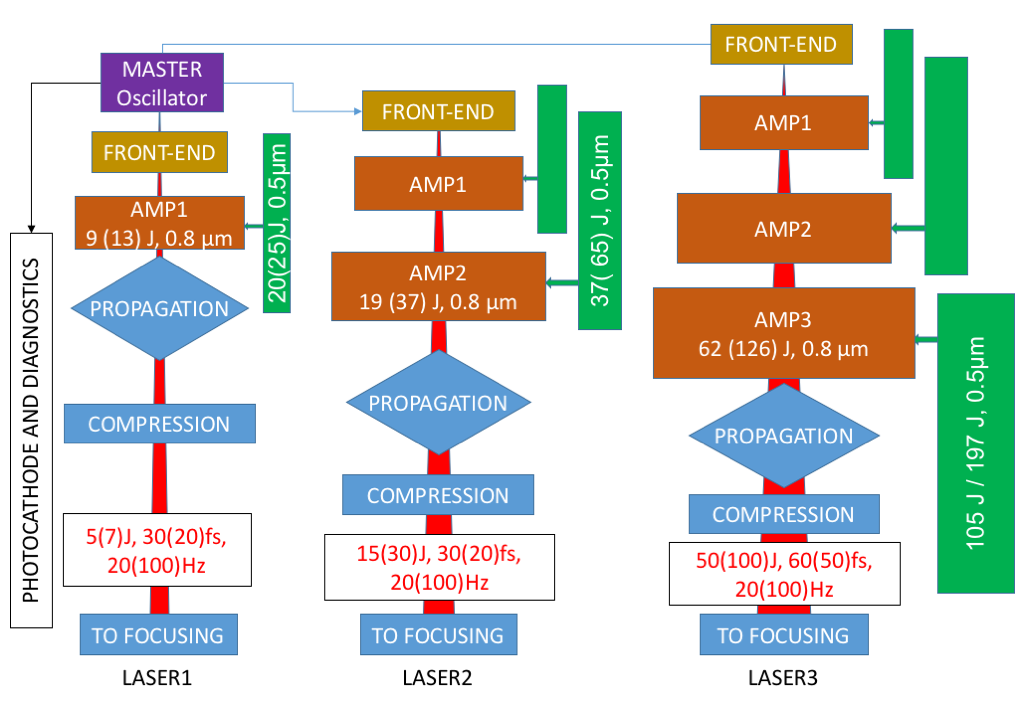}\caption{General layout of the EuPRAXIA laser, including the main blocks, from the common master oscillator, to the front-end, the power amplifiers, the propagation and the compressor. The main specifications of the system before and after the compression stage are also specified. The main pumping units are also included with the respective specifications at 0.5 $\mu$m wavelength.\label{fig:layout}}
\end{figure}

The baseline architecture of the EuPRAXIA laser relies on Ti:Sa technology and includes the most advanced components implemented so far in industrial systems plus leading edge technologies to conceive operation of the demanding detailed specification. Simultaneous operation of the three main laser systems and auxiliary laser beams for diagnostics and photocathode laser are considered, with a common oscillator. As shown in Figure \ref{fig:layout}, each laser chain includes a front-end, amplification sections, a propagation area and a compressor, with final transport to the target. 

As for the front-end, in order for the spectral amplitude/phase to be adjusted independently, three separated systems are envisaged, sharing the same oscillator and a common architecture. Each of these front-end will deliver stretched pulses with ~1J energy to the subsequent amplification stages.  

\subsection{Front-end}
The architecture adopted here is similar to several state of the art front-end systems available also as a commercial solution [Amplitude, 100 Hz]. After a common oscillator, shared among the three systems, a first CPA stage is devoted to the amplification of the pulse at the ~1mJ level, which is required to efficiently pump an XPW stage immediately downstream. The resulting pulse, with typical energy of the order of 10-100 $\mu$J is then seeded into the main CPA chain. The energy is then increased up to the >~1J level, needed by the subsequent amplification stages, by means of a regenerative amplifier and two multipass amplifiers, all running at 20Hz (P0 level) or 100Hz (P1 level). 

\subsection{Power amplification}
The amplification stages of the different beamlines will rely on Chirped Pulse Amplification in Ti:Sapphire, pumped by frequency doubled solid state Nd or Yb based lasers with emission in the range 515-532 nm. This technology is already well established for the amplification of ultrashort pulses down to a duration of <20 fs, and with energies >100 J. The most challenging point in the design of the required system is mainly related to the high pulse repetition rate, resulting in high average pump power requirements and thus setting a severe thermal load on the amplification stages. Indeed, petawatt lasers systems typically run in single shot mode or up to 1 Hz, whereas the EuPRAXIA uses a pulse repetition rate from 20 to 100 Hz.

A preliminary design of the amplification stages was  carried out considering the cooling requirements as one of the main and most demanding design drivers. Combination of conductive cooling through the crystal mounts and convective cooling in air of the amplifying crystal, the most commonly used cooling techniques in low repetition rate systems, may result insufficient  for high repetition rate operation at the required energy levels. In order to minimize pump power requirements and thermal dissipation needs, a detailed analysis was carried out of all the factors affecting the conversion from pump energy to energy on target, including extraction efficiency, transport and compression. 

Another driver in the design study was the modularity of the laser system. As pointed out before, some different plasma injection and amplification configurations are currently under consideration, resulting in a set of laser amplification chain systems with different requirements. These laser systems can be considered as consisting of a limited set of base modules that can be used alone, or built in multi-stage amplification chains, with minor adaptations. In view of an industrial development, this approach will help to reduce costs and development time.

\begin{figure}
\centering{}\includegraphics[scale=0.24]{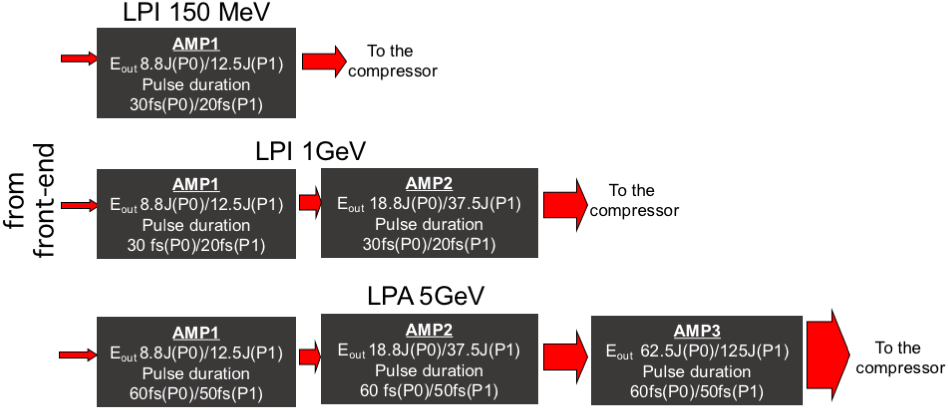}\caption{Layout of the power amplification sections of the three laser lines.
\label{fig:amps_layout}}
\end{figure}

A schematic layout of power amplification sections for each laser system is given in Figure \ref{fig:amps_layout}. Here the modules AMP1 and AMP2  have the same structure and they operate on the same input/output and pump energies, although they operate on pulses with different bandwidth, due to the different compressed pulse duration. The module AMP3 is the final energy booster specific for the 5 GeV accelerator.

Water cooling at room temperature of thin disk shaped crystals may be a
suitable solution. This approach has been recently proposed
\cite{Nagamihaly_2017} to achieve high repetition rate
operation in multi-PW Ti:Sapphire laser systems, and it is currently under
development in the frame of the ELI-ALPS related activities. Another possible
approach is the cooling of the crystals (shaped as thin slabs) with a high
speed gas flow at cryogenic temperatures, as used for instance in the DIPOLE
system (see for instance \cite{Banerjee_2012,Banerjee_2016}). This technology is currently demonstrated up to 10 Hz and an average pump power of 4.65 kW. With respect to water cooling this approach is more complex, and its cooling capability are limited by the smaller heat capacity of the flowing gas with respect to water.

Finally, the scalability of the technical solutions has also been considered as a guideline in the preliminary design study. The EuPRAXIA specifications for the various laser systems currently foresees two performance levels. The first one (P0) is considered as a (relatively) conservative performance, achievable with reasonable incremental developments of currently available laser technologies; the second one (P1) is more challenging, and it will require a more considerable technological effort. 

Based on these considerations power amplifiers were designed with attention to the individual modules and the respective dimensioning, in terms of pump energy, size and doping of the crystals, beam cross sections. Output energies, output pulse width, output beam profile, sensitivity to parameters, requirements on input pulse/beam were calculated using numerical simulations. Finally, cooling requirements and water cooling architectures have been investigated and a preliminary assessment of the thermal behaviour of the different modules was made including thermal aberrations.

\subsection{Module dimensioning}
\begin{figure}
\centering{}\includegraphics[scale=0.32]{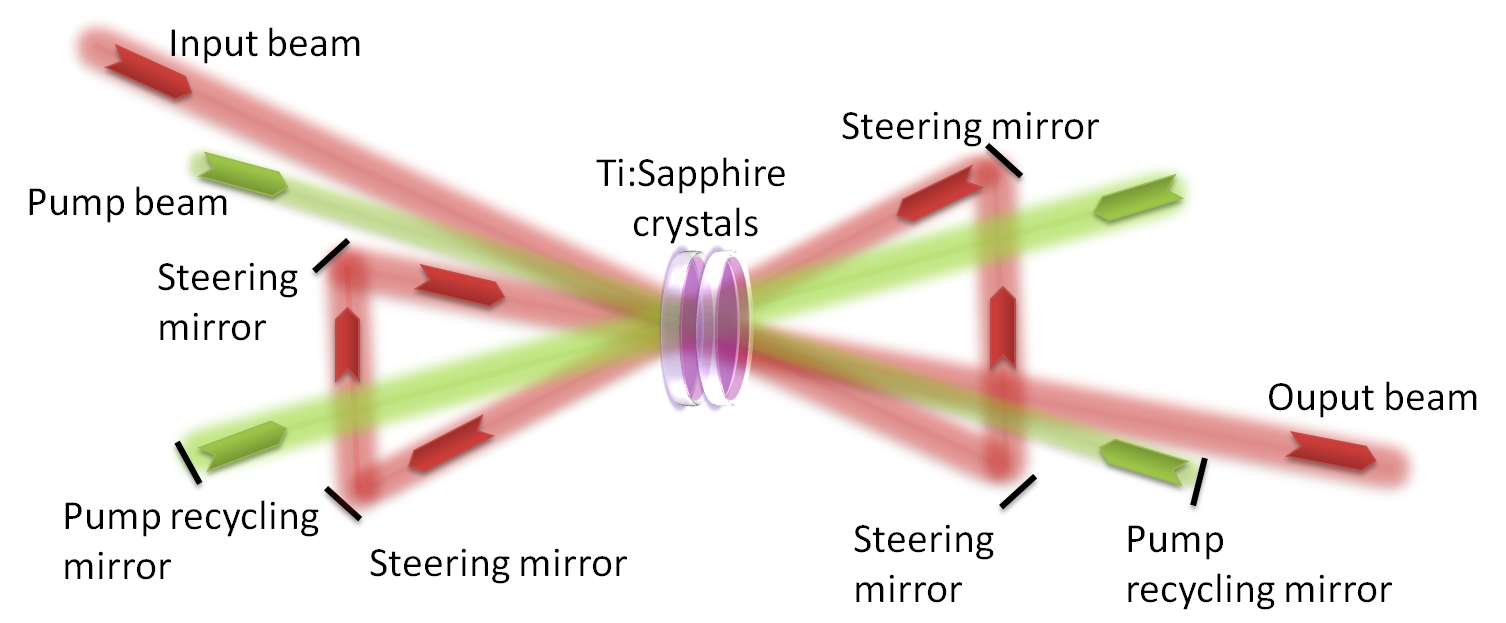}\caption{Schematic representation of a 3-pass amplification stage, in transmission geometry. The two pump beams are counter-propagating in the crystal, and after the first pass each one is back-reflected in the crystal for a second pass. The gain medium is considered to be split in two identical crystals.
\label{fig:multipass_layout}}
\end{figure}

The three laser modules described above were dimensioned by means of numerical
simulations using the code MIRO developed by CEA \cite{Morice_2003}, using the built-in amplification model for Ti:Sapphire. The simulations
were validated by comparison with real laser systems \cite{Gizzi2017,Chu_OE2013,Chu_LPL2013}. These case studies were considered relevant because the operating parameters of these systems are similar in terms of pump fluence, single pass gain and output energies. As a baseline design, all the amplifying modules feature a multi-pass amplification architecture, between 4 to 6 passes, depending on the stage, and a two-side injection of the pump beams. A conceptual scheme is shown in Figure \ref{fig:multipass_layout}, for a 4 pass layout.

As it will be described in details below, for thermal management reasons, the amplifying crystal will not be monolithic, but it will be divided in few identical sub-crystals (typically 2), face cooled by a liquid flow, to increase the available cooling surface. This fractioning of the crystal has no impact on the overall energy amplification performances.

In all the simulations, the seed pulse was assumed to have a Gaussian spectral shape, with a bandwidth corresponding to the desired pulse duration FWHM (assumed at the Fourier limit) with a stretched duration of 500 ps FWHM and a Gaussian temporal shape, regardless its bandwidth and compressed pulse duration, as a result of the stretching before the amplification.
The pump beam and the injected beam were assumed to have a super-Gaussian power distribution profile with index 8; pump and seed beam diameters were chosen so that the seed beam is slightly narrower than the pump beam, in order to have a more uniform gain profile. Each pump beam is assumed to pass twice in the amplifying crystal.

The amplifying crystals were considered cylindrical, with a clear aperture
larger than the pump beam of about 10 mm and a standard orientation, that is
with the c axis perpendicular to the propagation direction and parallel to the
polarization of the amplified beam. Doping level of the crystals has been
chosen to have a high absorption (>90\%) of the pump pulses, on a double pass
configuration. The absorption data with respect to doping level were derived
from commercial materials datasheets (see for instance \cite{Althecna_2017} [GT 2013]).
Pumping wavelength was assumed 532 nm. In case of different pumping wavelength (e.g. 514 nm or 527 nm) the pumping energy levels must be scaled with the ratio (532nm)/$\lambda_{Pump}$, to ensure the same pump photon flux. The doping level of the crystals must be revised accordingly.
Passive losses (e.g. at the optical interfaces, on the mirrors or due to volume scattering) were not considered in this phase.

The transverse gain $G_T$ was estimated on the basis of the pump fluence,
crystal size and doping using analytical formulas (see for instance
\cite{Chu_OE2013,Ertel_2008}). It is assumed
that the crystal lateral faces will be surrounded by an absorbing, index
matching liquid, or by an absorbing coating to prevent the onset of parasitic
lasing (PL) effects. Considering that the residual diffuse reflection
coefficient achievable at the crystal surfaces by these means is of the order
of $10^{-3}$ \cite{Chu_LPL2013} an estimated threshold level for
the suppression of the PL effects is $G_T<<10^{3}$. Moreover, to reduce the
transverse gain the amplifying crystals were pumped from both sides. To
mitigate parasitic lasing, Extraction During Pumping (EDP) was
implemented. Typically, a two-step pumping was adopted, with different pump
energy repartition between the two pumping steps. The transverse gain was
calculated for each passage, on the basis of the residual stored energy and
the injected energy, by means of the Frantz-Nodvick formulas \cite{Frantz_1963}. The sensitivity of the output energy to the pump and seed pulse energy was determined by calculating the variation output energy obtained with small variations in the pump and seed  energy; the relevant slopes were then calculated by linear interpolation. These slopes will be denoted as $S_{seed}$ and $S_{pump}$ below.

As an example we focus here on the output of the amplifier module AMP3 which is the most demanding in terms of size and specifications. The AMP3 module is based on a 6 pass amplification scheme, with re-pumping before the 3rd pass as EDP scheme. It is assumed that this stage is injected by the whole output energy available from AMP2. A summary of the output parameters is given in Table \ref{fig:output_amp3}. As it can be seen, the calculated output energy (62.4 J) matches the requirements.

\begin{table}
\centering{}\includegraphics[scale=0.35]{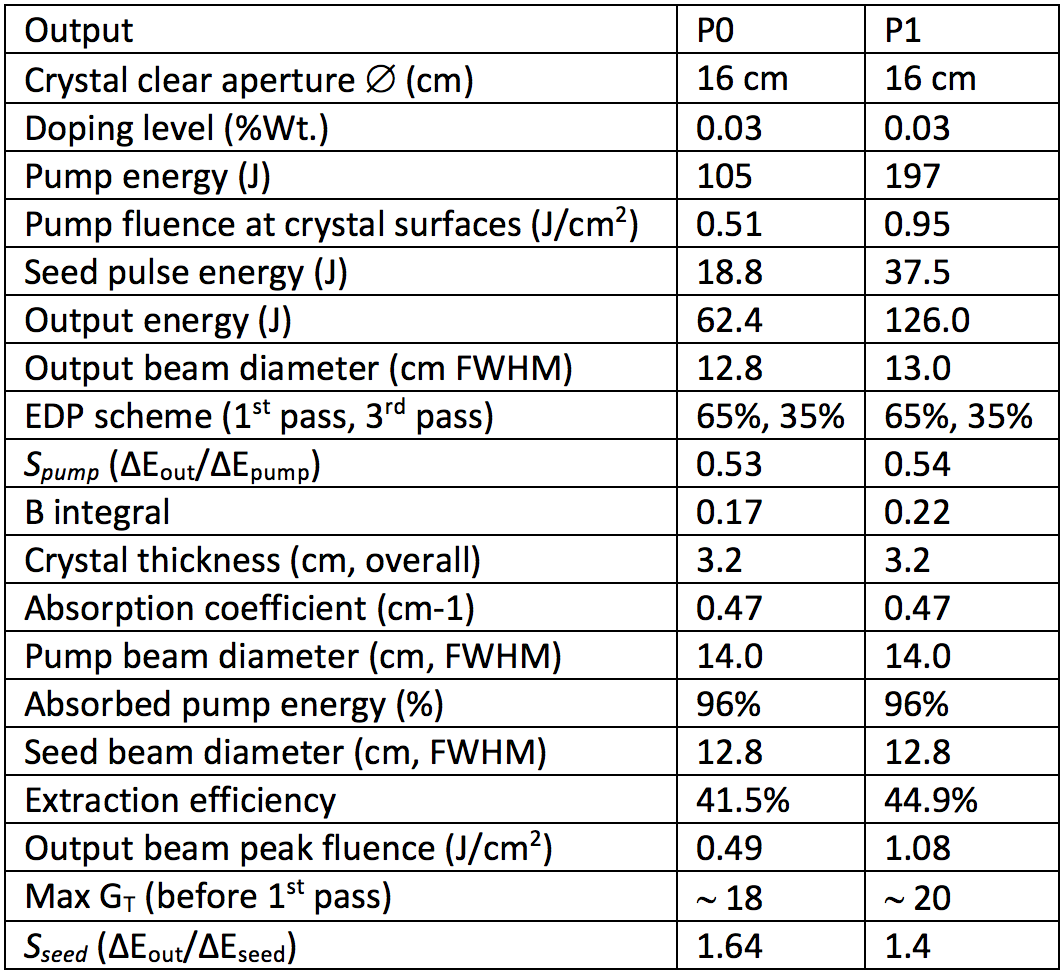}\caption{Main output parameters of the AMP3 module for the P0 and P1 level of performances as obtained from numerical simulations.
\label{fig:output_amp3}}
\end{table}

A set of output plots is shown in Figure {\ref{fig:plotsPass} including the pulse energy as a function of the number of passes, the input and output spectra and beam fluence profiles, for the P0 performance (top) and P1 performance (bottom). In the case of P0 performance level, due to the relatively long output pulse duration requirements (compressed output pulse duration 60 fs FWHM) the input pulse bandwidth is 18.7 nm FWHM, corresponding to a compressed pulse duration (to the Fourier Transform limit) of 50 fs. This leaves 10 fs of margin for incomplete recompression. The red shift of the amplified pulse peak was found 1 nm with respect to the input pulse, so that the input pulse spectrum must be peaked at 799 nm to have an output spectrum peaked at 800 nm. The spectral bandwidth at the output is 18.8 nm, with a negligible broadening with respect to the input pulse as shown in Figure \ref{fig:plotsPass}. Regarding the output beam shape, the FWHM of the output beam remains substantially unchanged with respect to the input beam as it can be seen in the top-right plot of Figure {\ref{fig:plotsPass}.
%\begin{figure}
%\centering{}
%\includegraphics[scale=0.15]{passes_p0} \includegraphics[scale=0.15]{spectrum_p0} \includegraphics[scale=0.15]{profile_p0}
%\includegraphics[scale=0.14]{passes_p1} \includegraphics[scale=0.15]{spectrum_p1} \includegraphics[scale=0.15]{profile_p1}

\begin{figure}
\centering{}
\includegraphics[scale=0.20]{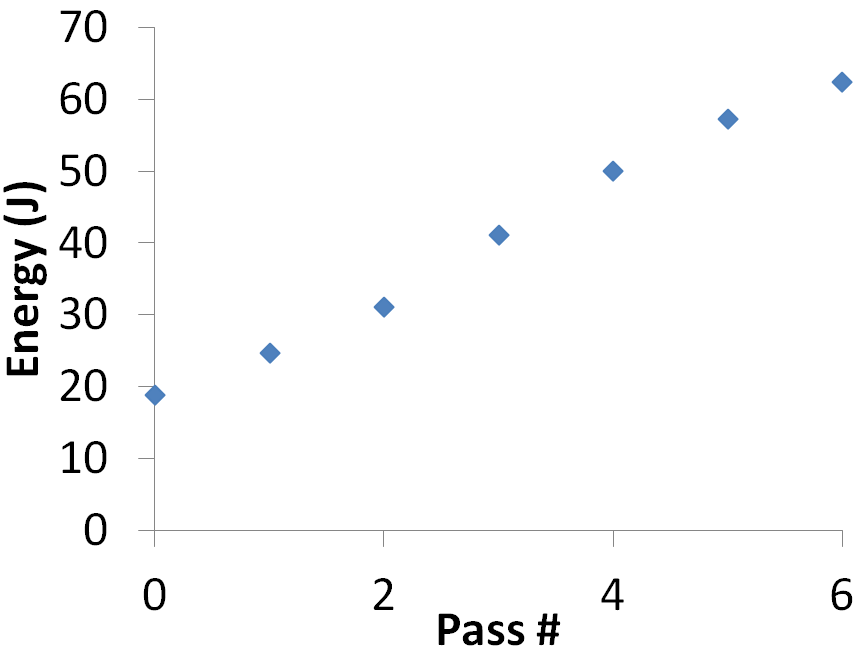}
\includegraphics[scale=0.20]{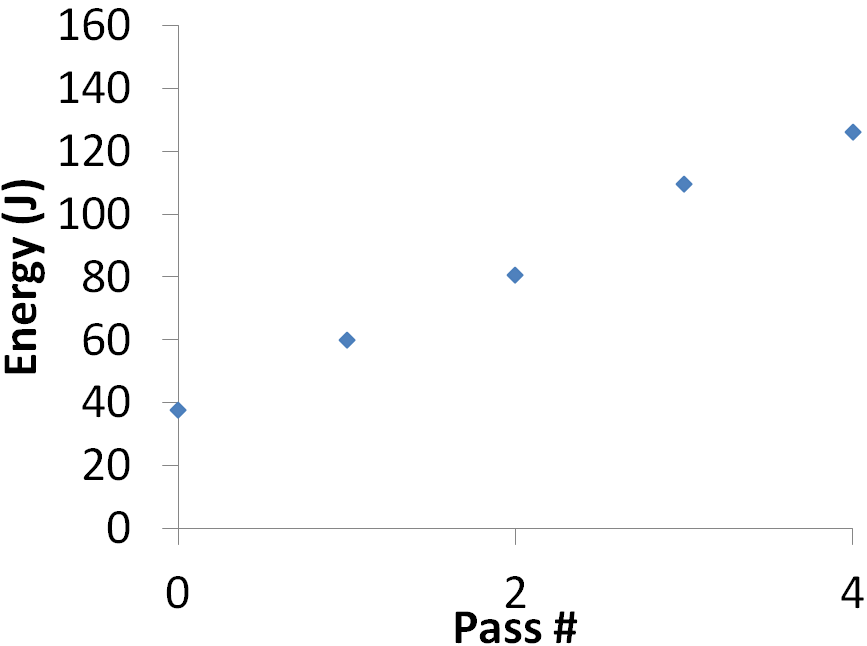}
\includegraphics[scale=0.18]{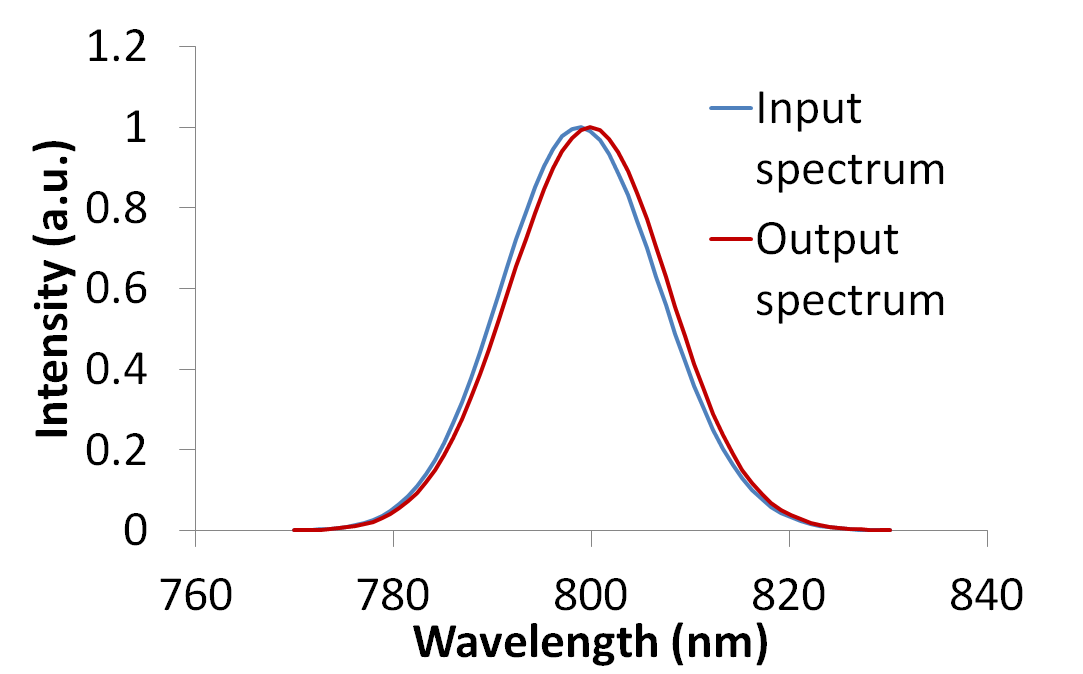}
\includegraphics[scale=0.18]{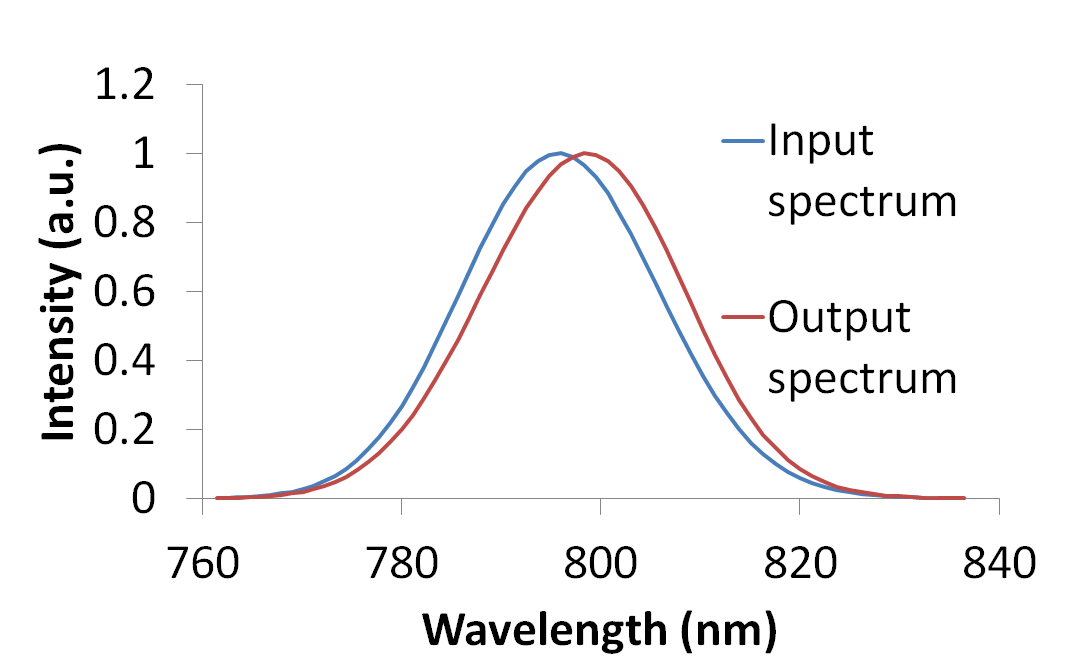}
\includegraphics[scale=0.18]{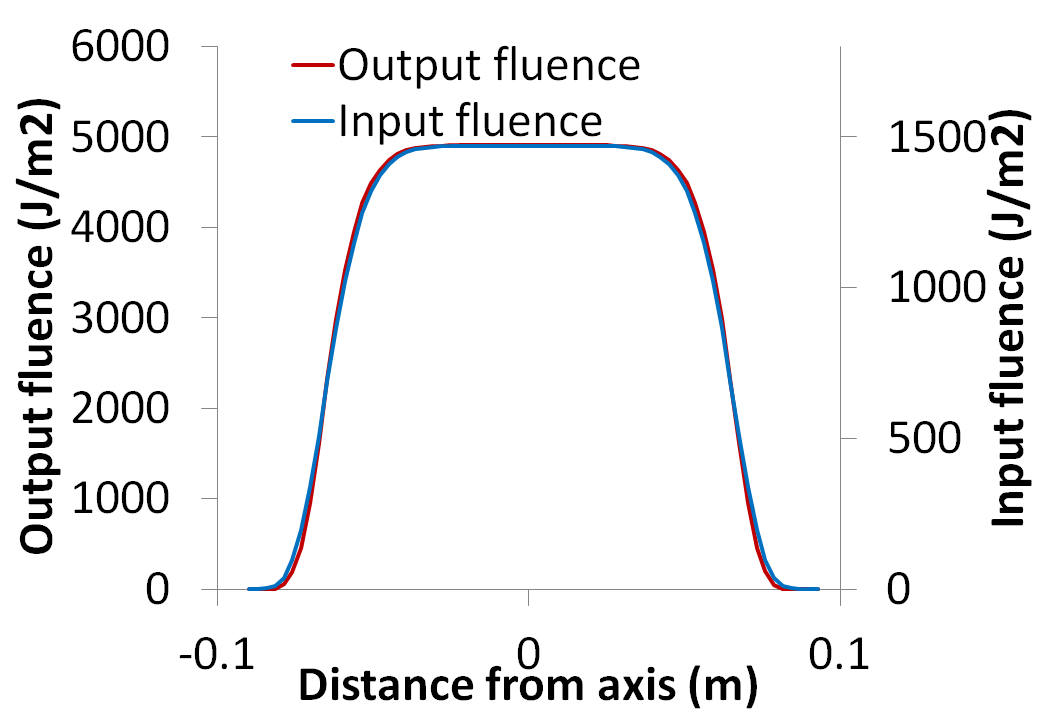}
\includegraphics[scale=0.18]{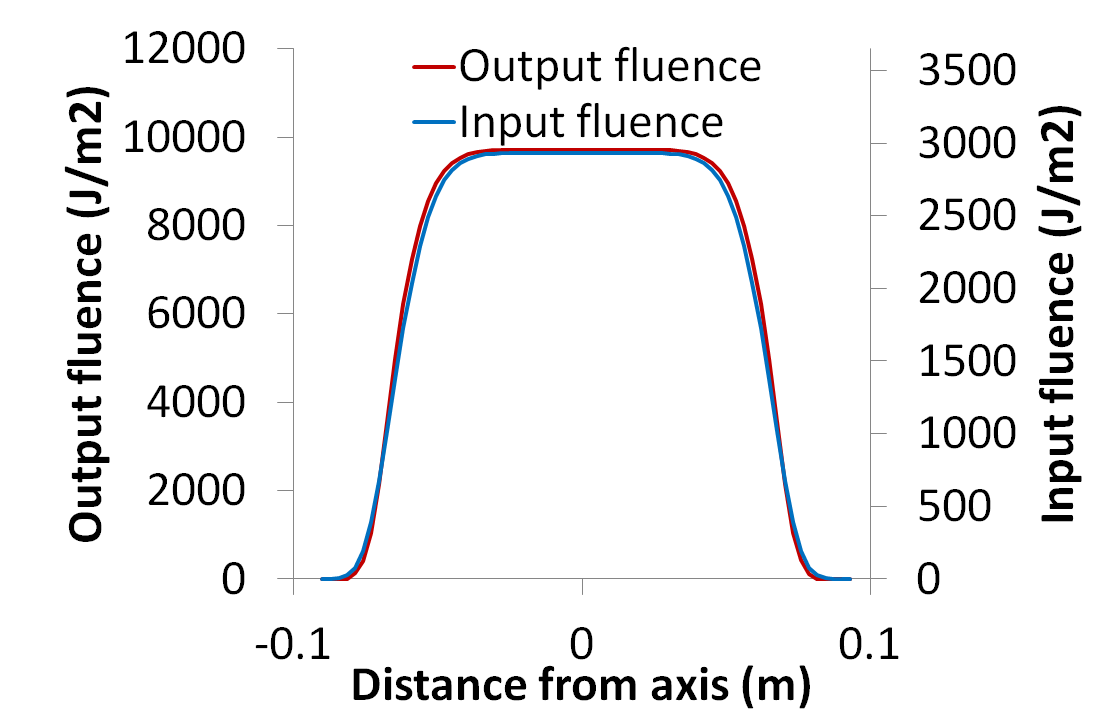}
\caption{Pulse energy as a function of the number of passes for the AMP3 module (top);  Input and output spectrum for the AMP3 module(middle);  Input and output beam fluence profile for the AMP3 module at P0 performance level (bottom). The two pro-files have been scaled to have the same graphical appearance. Note that the horizontal scale is in m and the vertical scale is in J/m$^2$ (=$10^4$ J/cm$^2$). The top set refers to the P0 performance level while the bottom set refers to the P1 performance level.
\label{fig:plotsPass}}
\end{figure}

For the P1 performance, the upscaling of the output performances is achieved by increasing the pump pulse energy, but leaving unchanged the geometrical parameters. The number of passes is reduced to 4 due to the higher energy of injection. The main parameters are resumed in the right column of Table \ref{fig:output_amp3}. As it can be seen, the calculated output energy (126 J) matches the required level as given in Table \ref{fig:laser_specs}. The pulse energy as a function of the number of passes is shown in the bottom-left plot of Figure {\ref{fig:plotsPass}. Due to the higher pump fluence and energy, the transverse gain is higher as well as the B integral, but both values are well within safety limits. The pulse bandwidth is larger than in the case of the P0 specification, because the required compressed pulse duration is smaller (50 fs FWHM). In this case a input pulse bandwidth of 23.3 nm FWHM, corresponds to a compressed pulse duration (to the Fourier Transform limit) of 40 fs. This leaves 10 fs of margin for incomplete recompression. The red shift of the amplified pulse peak was found 3 nm with respect to the input pulse, so that the input pulse spectrum must be peaked at 797 nm to have an output spectrum peaked at 800 nm. The spectral bandwidth remains substantially unchanged as shown in the bottom-centre plot of Figure {\ref{fig:plotsPass}. Regarding the output beam shape, the FWHM diameter of the output beam increases to 12.4 cm, due to the amplification of the beam wings, as it can be seen in the bottom-right plot of Figure {\ref{fig:plotsPass}.

\section{Pump lasers}
As outlined above, pump laser requirements are a critical aspect of our design. Our approach was to minimize pump laser needs while optimizing extraction efficiency and relying on efficient transport and compressor throughput and optimized second harmonic conversion efficiency of pump lasers.
The compressor throughput was assumed to be 80\%, achievable with a single grating reflectivity better that 95\%,. Such a value is not far from current commercial grating technology providing gold coated gratings with demonstrated reflectivity between 90\% and 94\% at 800nm. We also assume a conversion efficiency from 1$\mu$m pump energy to 0.5$\mu$m of 70\%, a conservative value if compared to the 80\% conversion efficiency demonstrated recently [De Vido, 2016]. Based on these assumptions, we obtain average IR pump power requirements ranging from 0.5 kW for the first amplifier module AMP1 at 20 Hz, to approximately 30 kW for the last amplifier module AMP3 at 100 Hz. The corresponding total pulse energies at 0.5$\mu$m range from 20 J to 200 J, resulting in a IR pump pulse energy ranging from 27 J to 280J. 
The survey of suitable pump lasers capable of delivering these performances was based on an extensive evaluation of currently available technologies.  Moreover, a similar analysis was carried out in the framework of the k-BELLA project (W.Leemans, EAAC 2017). These investigations consistently suggest that diode pumped Yb based systems are emerging as candidates to sustain the envisaged average power. At the same time, Nd based systems also proved to be capable of rep-rated operation at average power levels relevant for our aims. At this stage we can identify three different  technologies for the three EuPRAXIA lasers, all based on diode pumping or adaptable to diode pumping and all having been demonstrated at a sufficient average power level to give confidence on their technical feasibility.

\section{Thermal management}
The levels of thermal load considered here range for 200 W for the smaller energy amplifier module to 10kW for the maximum laser energy (AMP3) at the highest performance level (P1). We envisage two cooling techniques for the Ti:Sapphire crystal, namely cooling with water flow and cooling with He flow at cryogenic temperatures.  Water cooling of thin Ti:Sapphire crystal disks has been recently tested on a small scale system [Cvykow 2016] and  scaling up to kW levels have been studied Nagymihaly et al. [Nagymihaly 2017]. Thin disk (TD) technology may offer the possibility for Ti:Sa crystals to be used in high average output power systems because the longitudinal direction of heat extraction greatly reduces thermal lensing; scalability can be obtained by segmenting the required crystal length in thinner slices, each one individually cooled. The general scheme of this approach is shown in Figure 6.9-1. The surface of the disks is cooled by a high speed flow (several m/sec) of water at room temperature. The water flow is confined between the crystal surface and an optical window. The crystal can be cooled from both sides: in this case the use of counterpropagating water flows ensures a more uniform temperature profile across the disk aperture.

Cooling with gas flow at cryogenic temperatures is a technique currently under
advanced development for high average power, diode pumped Yb:YAG lasers, using
a multi-slab architecture. This technique has been adopted for the realization
of DiPOLE Yb:YAG amplifiers, targeting the generation of 10 J and 100 J ns
laser pulses at a repetition rate of 10 Hz
\cite{Banerjee_2012,Banerjee_2016,Mason_2015}. It has also be applied to the development of an Yb:CaF2 high
energy amplifier \cite{Siebold_2013}. Cooling is obtained by a He gas flow with temperature in the range 125-175 K, 5-10 bar of pressure and a flow velocity around 25 m/sec.
For several reasons, water cooling of thin disks is considered as the candidate technology for the cooling of the amplification stages. In fact, water cooling at near room temperature is easier to implement than gas cooling at cryogenic temperature and the cooling capability achievable with the water flow is higher than with the gas flow, due to the much higher heat capacity of the cooling fluid. Indeed, heat removal rates per unit surface, expressed by the average Heat Transfer Coefficient, HTC) are about 1 order of magnitude higher for water cooling than for gas cooling; this is particularly relevant in view of the scaling up of the system toward high average power levels.
In order to achieve a sufficient heat removal rate, gas cooling requires a large heat exchange surface. This implies that the gain medium is usually split in several slabs, increasing the occurrence of parasitic reflections at interfaces, which can be detrimental for the pulse contrast. Conversely, for a given overall gain length water cooling allows using a smaller number of thicker slabs thus reducing the risk of parasitic reflections. A potential drawback of the water cooling technique is that the turbulences and temperature gradients in the water flow can induce aberrations in the laser beams. This can be critical in particular in the schemes where the amplified beam crosses the cooling flow. 

\begin{figure}
\centering{}\includegraphics[scale=0.65]{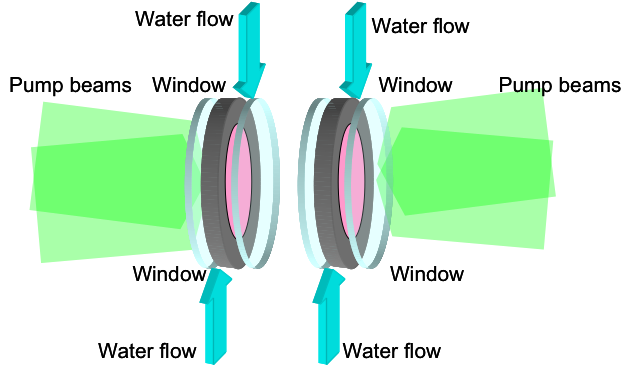}\caption{Schematics of the twin thin disk crystal assembly. Each crystal is individually cooled by two counterpropagating water flows, confined by an optical window. The outer ring of the crystal features an absorbing region or edge coating for ASE and parasitic lasing suppression. Pump beams are also represented.
\label{fig:cooling_scheme}}
\end{figure}

Nonetheless, a similar approach has already been experimented in some high power laser sources. For instance Fu et al. \cite{Fu:14} have recently proposed a 3 kW CW liquid cooled Nd:YAG laser. Multiple Nd:YAG slabs are face cooled by a flow of deuterated water (D2O) which is crossed by the laser beam. Here D2O is used instead of H2O because of the absorption at 1064 nm of H2O, but it should not be needed at 800 nm. A similar cooling scheme was adopted by Wang et al. \cite{Wang:16} for a 7 kW Nd:YAG laser system. Also in this case the gain medium is split in several slabs, face cooled by a high speed D2O flow crossed by the laser beam. It must be noticed that both devices are oscillators and not multi-pass amplifiers. This is an even more critical arrangement because the effect of the aberrations induced by the flow turbulences are accumulated over the high number of intra-cavity passes. Finally, a similar approach was demonstrated to be effective in the so-called Disk Amplifier Head (DAH) used in the P-series by Amplitude Technology.
A possible alternative, that was also considered in the design, is the use of a reflection geometry where the laser beam is reflected on the back side of the amplifying crystal. In this case the crystal is fluid cooled on its back side only, so that the laser beam does not cross the cooling fluid flow. This eliminates the problem of the fluid turbulences affecting the propagation of the laser beam, at the price of a reduction of the available cooling surface. This geometry was experimented for instance by V. Cvhykov et al. \cite{Chvykov:16} for a Ti:Sapphire amplifier.

In the case of the transmission scheme with two crystals with half-length is depicted in Figure \ref{fig:cooling_scheme} the heat removal capability of the system is doubled with respect to a single crystal and a relatively low crystal doping level can be used, easing parasitic lasing suppression. 

For the preliminary evaluation of thermal effects, it was assumed that about 50\% of the absorbed pump energy is dissipated as heat in the crystals (heat efficiency factor =0.5). The thermomechanical behavior of the crystals was modeled using a commercial software package, i.e. LAS-CAD (ver. 3.6.1) developed by LAS-CAD GmbH (www.las-cad.com). The software features several numerical tools for the modeling of solid state laser systems. Among them, a Finite Element Analysis (FEA) module was extensively used for the purposes of this preliminary design. The FEA thermal modeling was used to calculate the spatial temperature distribution in the gain material resulting from the heat input due to the pump absorption and from the cooling at the surfaces. The stress mechanical modeling was then used to calculate the stress distribution and the deformation distribution in the crystal, induced by the thermal expansion. 

The thermal aberrations were calculated by computing the Optical Path
Difference (OPD) distribution across the crystal aperture. Total OPD is made
of two main contributions, namely the variation in the optical path length due
to the variation of the refractive index with temperature integrated along the
crystal length L and the variation in the crystal thickness due to thermal
expansion and thermally induced stresses. Thermally induced birefringence \cite{Ferrara_2014} was not considered and the dependence of the thermal and mechanical parameters of the Ti:Sapphire from the temperature was neglected, as well as the slight anisotropy of some parameters in the orientation with respect to the crystalline axes. 

For the thermal modeling of the AMP3 module, the crystal was assumed to have an overall diameter of 18 cm, with a cooled surface diameter of 16 cm. The absorption coefficient for the pump radiation was assumed to be 0.47 cm$^{-1}$. 

\begin{figure}
\centering{}\includegraphics[scale=0.50]{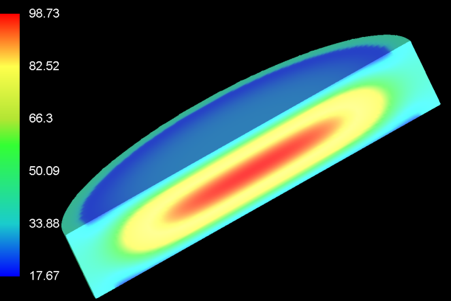}
\centering{}\includegraphics[scale=0.50]{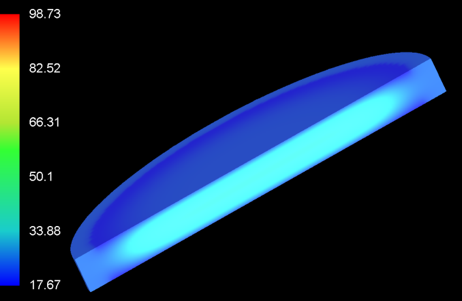}

\caption{Temperature distribution (in °C) at the performance level P1, for a single crystal amplifier (left) with thickness 3.2 cm and for one of the two crystals with thickness 1.6 cm (right). Diameter of the crystal is 18 cm in both cases.
\label{fig:temperature_distr}}
\end{figure}

The plots of Figure \ref{fig:temperature_distr} shows the average temperature distribution for the P1 levels of performance for the single slab and for the thin slab with half of the thickness. In the single crystal configuration, a peak temperature of about 100 °C was found, which is probably too high, with a surface temperature around 40 °C. On the other hand, the split (thin) crystal configuration features a much lower peak temperature of only 40.8 °C that has a positive impact in terms of overall thermal aberrations.   The analysis of the thermal aberrations shows that the maximum OPD for the thin crystal is 7 $\mu$m, a factor of 3 smaller than in the case of the thick crystal. Similarly, the dioptric power of the equivalent spherical thermal for the thin crystal is 2.7E-3, i.e. approximately 30\% of the thick crystal case. In general it appears that water cooling at high flow speed is a promising approach for the thermal management of the amplification stages, provided a cooling scheme based on the splitting of the crystal in twin gain elements is adopted to mitigate thermal aberrations.

\section{Candidate pumping units}
Several systems are emerging in the current scenario of diode-pumped solid state lasers for high average power. Among these, three configurations have demonstrated performances relevant to our specifications. 
Based on an end-pumped stack of ceramic Yb:YAG slabs, the DiPOLE system is a diode-pumped, solid state laser amplifier architecture cooled by a flow of
low-temperature, high-pressure helium gas [Ertel, 2011].  This technology was recently demonstrated at the 1kW level, showing 100 J output energy at 10 Hz,
1030 nm with > 60 J expected conversion @ 515 nm \cite{Mason_2015}.  This architecture and gain material exhibit reduced reabsorption loss and increased absorption and emission cross-sections in Yb:YAG, with a low quantum defect due to the very close pump and emission wavelengths being 940 nm and 1030 nm,
respectively. These advantages enable efficient energy extraction and potential scalability to high average power.  ceramic Yb:YAG, ceramic Nd:YAG
and glass Nd.YAG. Several laser systems based on this amplifier technology have already been constructed and successfully operated. These lasers produce
pulses of ns duration, 1030 nm wavelength, multi-J energy and a repetition rate of 10 Hz. Frequency doubling to 515nm, required for pumping of
Ti:sapphire amplifiers, has also been successfully demonstrated. Performance demonstrated so far includes generation of 105 J, 10 ns, 1030 nm pulses at
10 Hz  and frequency doubling of 7 J pulses at 10 Hz with 82 \% conversion efficiency. Recent progress \cite{DeVido_OME2017} also shows the possibility of Increasing the output energy of an existing system to 150 J at 1030nm and demonstrating frequency doubling at the 100 J level, with an expected output > 100 J at 515 nm. 
Based on these considerations, DiPOLE-like architecture could provide the necessary building blocks for pump lasers up to Laser2 at P0 with some moderate energy scaling and staggering of two 10 Hz units to match the required 20 Hz repetition rate. 
Delivering pump pulses at 100 Hz at the > 100 J energy level using the current DiPOLE amplifier design is likely to need new approaches, such as liquid cooling and room temperature operation. These new approaches, whilst requiring substantial up-front investment, have the potential to significantly reduce the number of lasers, and therefore the cost, required for a given amount of average power.

Using Ceramic Nd:YAG, the P60 is a commercial system produced by Amplitude Technologies as a part of the P-series systems, using an improved "active mirror" configuration, the so-called Disk Amplifier Heads (DAH). Currently delivered with flash lamp pumping, conversion to diode pumping has been developed by Amplitude and a preliminary design has been produced.  The general architecture of the P60 includes a seeder, followed by 6 identical Nd:YAG DAH amplifiers. The main features of the system are 1) enhanced and simple thermal management without cryogenic cooling with longitudinal liquid cooling in gain/heat load distributed disks; 2) compatiblility with diode pumping, 3) compact footprint 1.5 x 4.8 m and beam specifications suitable for Ti:Sa pumping. The demonstrated performances include >70J at 1064nm with a 0.41\% RMS shot-to-shot stability and  75\% SHG efficiency with 5-6ns Gaussian temporal pulse @ 2.8J/cm$^2$ incident fluence with a 65 mm diameter and 18 mm thick LBO type I crystal. These values indicate that an array of a limited number of P60-like systems could deliver the required pump energy for Laser1 and Laser2. Laser3 will require additional development to reach the >100 J level after SHG so that an array of up to 4-5 units would provide sufficient pump energy. The dpssl version of the P60 is envisaged to operate at an ultimate rep. rate of 100 Hz. However, current heat load performances would already enable 3.5 kW average power with 75J per pulse (IR) at 50 Hz with no risk. In fact, the thermal load extraction capability demonstrated in disk amplifiers with flashlamps pumping is similar to the heat load at 50Hz with diode pumping. Moreover, diodes with required brilliance and Power supplies are existing and qualified.   One important issue concerns the lifetime of diodes for high rep rate.  Assuming a diode lifetime ~ 2 billions shots at 2\% duty cycle, diodes should be replaced every ~700 days at 8/24H operation at 100Hz rep-rate and every ~1400 days at 8/24H operation at 50Hz rep-rate.

Based on a diode pumped, He cooled Nd:glass with cooled ASE edge cladding, the HAPLS system developed by the LLNL for the ELI Beamlines can operate at  >100 J output energy demonstrated @ 3.3 Hz, 1053 nm, with 0.7\% RMS stability and 80 J SHG energy @ 526.5 nm. Ramping up to 10 Hz, 200 J (IR) design limit is currently in progress. In the HAPLS system, the diode pumped system is used to pump a Ti:Sa system to deliver 30 J in 30 fs at 10 Hz. This technology is derived from Inertial Fusion Energy Laser architectures and can be aperture scaled to single aperture, kilojoule, >100kW output.  Based on the data obtained from the design, construction, and operation of this laser system, is envisaged that increasing the repetition rate by 10 times to 100Hz is possible.

Summarising the current and expected performances of kW-scale DPSSL pumping systems currently available, we can conclude that the three selected systems are in a highly advanced development stage. All of them have demonstrated performances required to fulfil part, if not all of the pumping needs of the main EuPRAXIA lasers P0 Specifications. Some of these technologies are being considered independently for scaled operation at higher rep rate and higher energy per pulse, making them good candidate for pump lasers to reach the P1 specifications. However, a significant targeted development will be needed to address scaling of candidate systems at the desired 100 Hz repetition rate. 

\section{Output Energy Stability}
Energy fluctuations of both the input pulse and of the pump energy determine the occurrence of fluctuations in the output energy. Moreover, when the individual stages are assembled in an amplification chain, the fluctuations in the earlier stages influences the output energy. This effect tightens the requirements on the stability of the front-end and on the stability of the pump sources. 
The evaluation of the fluctuation of the output energy,  determined by the fluctuations in the pump and seed energy, was carried out on the basis of the results of the simulations described above.

\begin{figure}
\centering{}
\includegraphics[scale=0.5]{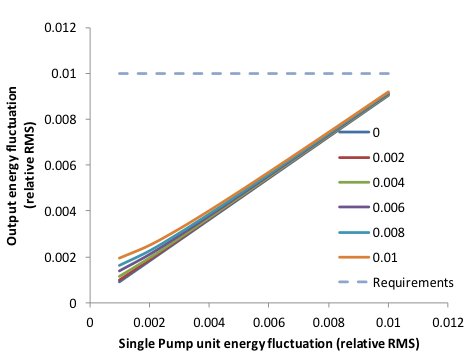} \includegraphics[scale=0.5]{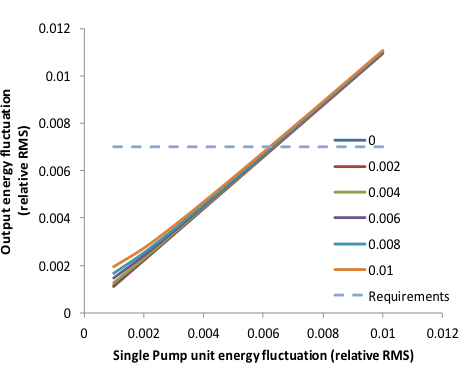} 
\caption{Relative variation of the output energy (relative RMS) for the Laser3, as a function of the relative fluctuations of the output energy of the individual pump unit. The different curves in the graphs correspond to different relative energy fluctuation levels (RMS) of the seed pulse generated by the front-end. Left graph correspond to the P0 performance level, right graph corresponds to the P1 performance level.
\label{fig:plots}}
\end{figure}
It was assumed that each different stage is pumped by one or more identical pump laser, with an energy output of about 60 J at 515-532 nm. This reflects the expected performances of possible pump lasers. Given an absolute energy fluctuation level for each individual pump source, assumed identical for all the modules in the pump laser array, and assuming again that the fluctuations of the individual sources are statistically uncorrelated, we find the overall absolute pump energy fluctuation as shown in Figure \ref{fig:plots} for the case of Laser3.
In general, the results of the analysis shown above indicate that i) the front-end energy fluctuations have a relatively small impact on the overall system stability, which is mainly influenced by the fluctuations of the pump sources; ii) as far as the P0 performance level is concerned, the requirement of the overall output stability <1\% allows a fluctuation of the individual pump source up to about 0.8\% (up to 1.1\% in the case of Laser3); iii)concerning the P1 performance level, a stability of the individual pump module of 0.6\% RMS should be sufficient to attain an overall stability level in the output energy of <0.7\%.

\section{Beam Transport Control}
Another very challenging task in the EuPRAXIA laser design is the transport to the interaction chamber after amplification and here we briefly discuss the main issues currently under scrutiny. The use  of the most effective technologies is foreseen, as shown in the conceptual diagram of Figure \ref{fig:transport},  including adaptive optics, spatial filters, cooled grating technology for the compressor and compact focusing optics.  
\begin{figure}[ht]
\centering{}
\includegraphics[scale=0.31]{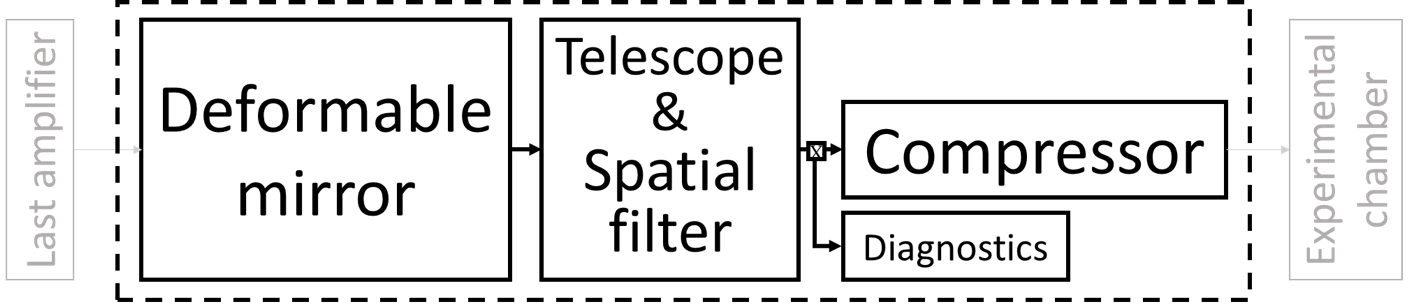}
\caption{Schematic view of the main beam transport components from the last amplifier to the interaction chamber.}
\label{fig:transport}
\end{figure}

The main challenges are certainly the thermal and spectral issues in the compressor, and the pointing stability at the interaction plane. As for the compression, EuPRAXIA needs a compressor able to handle around $10$ W/cm$^2$ in a $\sim 10^{-6}$ mbar vacuum. Regardless of the grating technology, cooling of the gratings will most likely be necessary. Moreover, depending on the laser beamline, wide bandwidth acceptance will be needed. In the current state of the art, studies about thermal issues on the gratings are only in their primordial stage. At LLNL [Alessi, 2016] measurements were carried out of surface deformations of gold gratings at hundreds of Watts average power. As a baseline, in EuPRAXIA we foresee the use of gold gratings, where a cooling strategy in vacuum is being studied in deep detail. Nevertheless, new grating technologies are emerging, like hybrid gratings with metal and multilayer dielectric coatings or new configurations for pure multilayer dielectric gratings, to increase the compressor global spectral acceptance and work at $800$ nm central wavelength. Research in this direction is very active, so  technology of these novel solutions will likely reach readiness for EuPRAXIA construction time scale.\\
Regarding pointing stability, in general, the requirements depend on the specific application involving superposition of two laser focal spots or a laser focal spot and an electron beam in the focal region, as in external injection or Thomson scattering \cite{Gizzi2009, 4599118}, or the matching of one or more laser focal spots with plasma targets \cite{doi:10.1063/1.4975839}.  From the point of view of intensity in a given position in the focal spot, specifications aim at less than $10\%$ for Laser1 and 2 and below $5 \%$ for Laser3, which call for high stability for the performance of the amplification chain as discussed above.  From the mechanical point of view, Laser3 is the most demanding requiring $1$ $\mu$rad (or less) pointing stability. These specifications will require a major design effort, still under investigation. In general, pointing stability will be affected by the whole system and several strategies can be considered. Active stabilisation of PW-scale lasers using closed loops between stability detection on the interaction point and crucial transport optics are being tested recently \cite{GALIMBERTI_CHANNELING}. At the same time, passive stabilisation acting on building specification and adopting self-stabilizing optics mounts, are also being considered.

\section{Conclusions}

The design of the EuPRAXIA laser driver is, by all means, a very challenging task. The required laser performances are unique in the current world wide scenario and dramatic developments are ongoing on several aspects, including architectures and components. Guided by the physics requirements and taking into account the expected time-scale of the construction of the EuPRAXIA infrastructure, we selected our enabling technology, define a preliminary architecture and conceive our initial layout. Our concept combines most, if not all, of the most advanced and proven technologies and aims at delivering performances well beyond those of existing systems or systems currently under construction.
With this design, EuPRAXIA will provide a credible step towards a reliable operation of a laser-driven plasma accelerator, giving opportunity to the EU laser industry to gain momentum and setting the basis for another step-change in high power laser based sciences and applications.

\section{Acknowledgements}
The   research   leading   to   these   results   has   received
funding from the EU Horizon 2020 Research
and   Innovation   Program   under   Grant   Agreement   No.
653782 EuPRAXIA. This project has received funding from the CNR funded Italian research
Network ELI-Italy. We gratefully
acknowledge P. Mason and K.Ertel of the Central Laser Facility, A.Bayramian, 
C. Siders and C. Haefner of the Lawrence Livermore National Laboratory, 
and F.Falcoz of Amplitude Technologies for fruitful discussion on the pump 
laser developments.

%\begin{thebibliography}{10}

\bibliography{citationsThisPaper}
\bibliographystyle{elsarticle-num}

%\bibitem{mason}Mason, P.; Divoky, M.; Ertel, K.; Pilar, J.; Butcher,
%T.; Hanus, M.; Banerjee, S.; Phillips J., Smith, J.; De Vido, M.;
%et al. Kilowatt average power 100 J-level diode pumped solid state
%laser. Optica 2017, 4, 438\textendash 439.
%\end{thebibliography}

\end{document}